\documentclass[aip,twocolumn,letterpaper,nofootinbib]{revtex4}%
\usepackage{graphicx}
\usepackage[colorlinks=true,linkcolor=black,urlcolor=blue,citecolor=blue]{hyperref}
\usepackage{upgreek}
\usepackage{titlesec}
\usepackage{mhchem}
\titlespacing\section{0pt}{6pt plus 2pt minus 2pt}{6pt plus 2pt minus 2pt}
\titlespacing\subsection{0pt}{6pt plus 2pt minus 2pt}{6pt plus 2pt minus 2pt}
\titlespacing\subsubsection{0pt}{6pt plus 2pt minus 2pt}{6pt plus 2pt minus 2pt}
\AtBeginDocument{%
 \abovedisplayskip=9pt
 \belowdisplayskip=9pt
}

\begin{document}

\title{A USB-controlled potentiostat/galvanostat for thin-film battery characterization}
\thanks{Supplementary material available online: \cite{sup}}
\author{Thomas Dobbelaere}
\email{thomas.dobbelaere@ugent.be}
\affiliation{Department of Solid State Sciences, Ghent University, Belgium} 
\date{\today}

\begin{abstract}
This paper describes the design of a low-cost USB-controlled potentiostat/galvanostat which can measure or apply potentials in the range of $\pm$8V, and measure or apply currents ranging from nanoamps to max.~$\pm$25~mA. Precision is excellent thanks to the on-board 20-bit D/A-convertor and 22-bit A/D-convertors. The dual control modes and its wide potential range make it especially suitable for battery characterization. As an example use case, measurements are presented on a lithium-ion test cell using thin-film anatase \ce{TiO2} as the working electrode. A cross-platform Python program may be used to run electrochemical experiments within an easy-to-use graphical user interface. Designed with an open hardware philosophy and using open-source tools, all the details of the project (including the schematic, PCB design, microcontroller firmware, and host computer software) are freely available, making custom modifications of the design straightforward.
\end{abstract}

\maketitle

\section{Introduction}

\subsection{Potentiostat basics}

\noindent The potentiostat is an essential tool in electrochemical research. It allows the experimenter to apply a potential to a system (i.e. an electrochemical cell) and measure the resulting current, or \emph{vice versa}. The unique property of the potentiostat -- the thing that differentiates it from a simple combination of an adjustable voltage source and an ammeter -- is that it can do so while keeping the path where the current flows separate from the path where the potential is measured. This is necessary in electrochemical cells because potentials are usually measured against a ``reference electrode'' which only provides an accurate and stable potential in equilibrium conditions, i.e.~when it is not disturbed by any current flow. Using two terminals for current flow and two others for potential sensing, a total of four electrode connections are needed; they are usually named as follows:
\begin{itemize}
\setlength\itemsep{0pt}
\item Working electrode (abbreviation: WE)
\item Counter electrode (abbreviation: CE)
\item Sense electrode (abbreviation: SE)
\item Reference electrode (abbreviation: RE)
\end{itemize}
In a four-electrode connection scheme, the potential is measured (and no current flows) between SE and RE, and current is applied (regardless of the voltage drop) between WE and CE.\footnote{It should be noted that the potentiostat is functionally equivalent to a source-measure unit with separate force/sense lines, with the WE/CE pair corresponding to the ``force'' connections and the SE/RE pair corresponding to the ``sense'' connections.}

\begin{figure}[h!]
 \centering
\includegraphics[scale=1.05]{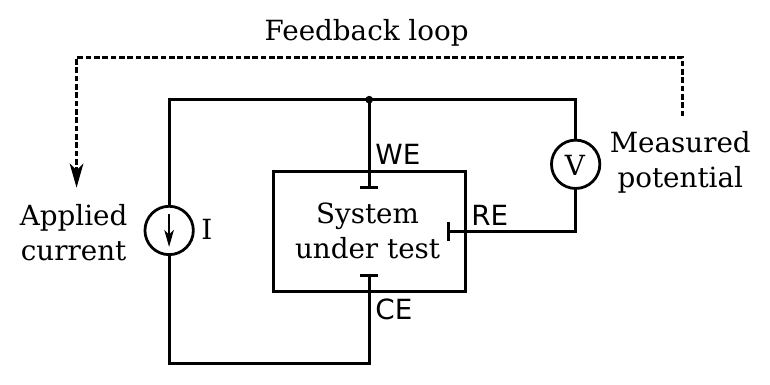}
  \caption{Schematic diagram showing the three-electrode connection scheme between the electrochemical cell and the potentiostat, resulting in separate paths for the application of current (between WE and CE) and the sensing of potential (between WE and RE).}
  \label{fig1}
\end{figure}

In most electrochemical cells, the SE and WE are tied together (and still referred to as WE), resulting in a three-electrode connection scheme: the potential is measured between WE and RE, and current flows between WE and CE. This is illustrated in Figure~\ref{fig1}. Using this scheme, one would naively assume that it is impossible to control the potential; as there is zero current flow, the potential can only be measured and not forced to a certain value.

The potentiostat can, however, apply current between WE and CE, which in turn influences the potential between WE and RE; thereby, using a feedback loop, any desired potential between WE and RE can be achieved by applying whatever current is necessary between WE and CE. This control mode is called the ``potentiostatic mode'': it allows the user to set a desired potential, and the potentiostat will try to reach that potential by adjusting the current.

\begin{figure*}[t]
 \centering
\includegraphics[scale=1.05]{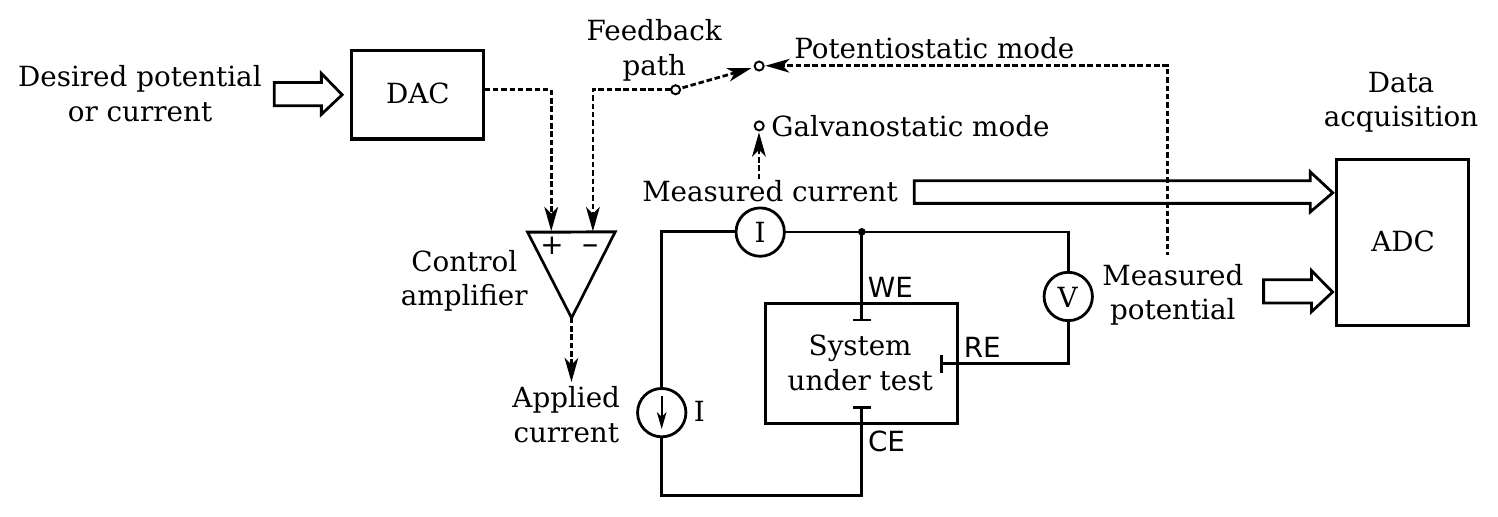}
  \caption{Block diagram illustrating the working principle of the potentiostat when it is controlling an electrochemical cell. The device is able to switch between potentiostatic (potential control) or galvanostatic (current control) modes by changing the feedback path.}
  \label{fig2}
\end{figure*}

Alternatively, it is often useful to have control over the current instead (i.e.~allowing the user to set a desired current, no matter what the resulting potential may be); this is called the ``galvanostatic mode''. The presented circuit elegantly combines both of these modes by making it possible to switch the feedback path between the potentiostatic and the galvanostatic modes, as shown in Figure~\ref{fig2}. On a basic level, the circuit operates as follows:

\begin{itemize}
\setlength\itemsep{3pt}
\item A digital-to-analog convertor (DAC) outputs an electrical signal representing either the desired potential (in the potentiostatic mode) or the desired current (in the galvanostatic mode).
\item An operational amplifier compares this to the measured potential (in the potentiostatic mode) or the measured current (in the galvanostatic mode), and drives current into the CE until the measured value equals the DAC setpoint.
\item Both the measured potential and the measured current are fed into an analog-to-digital convertor (ADC) for data acquisition purposes.
\end{itemize}

\subsection{Comparison to previously published designs}

Although a number of similar open source, ``do-it-yourself'' potentiostat designs have recently been published \cite{friedman2012,rowe2011,dryden2015}, these designs are not suitable for battery characterization. The Friedman et al. design \cite{friedman2012} does not allow the working electrode potential to be scanned (it can only be adjusted to a fixed value in hardware), limiting its use to chronoamperometry (i.e. recording the working electrode current as a function of time). The CheapStat \cite{rowe2011} supports a number of electrochemical techniques including cyclic voltammetry, but its potential range is limited to $\pm$1\,V. The DStat \cite{dryden2015} improves upon the CheapStat in several ways, and has impressive low-current capabilities, but it is still limited to a potential range of $\pm$1.5\,V. Although these potential ranges are wide enough for many aqueous electrochemistry experiments, they are insufficient for e.g.~lithium-ion batteries which can reach cell potentials over 4\,V. The potential range of the design described in this paper is $\pm$8\,V.

Another highly desirable feature for battery characterization is the inclusion of a galvanostatic mode. The aforementioned designs implement the ``adder potentiostat'' topology \cite{bard_faulkner} which only provides potentiostatic control. The presented design has a different topology which enables switching between potentiostatic control and galvanostatic control with a single (digitally controlled) switch.

\subsection{Comparison to commercial instruments}
While there are plenty of commercial instruments which can be bought from manufacturers such as Metrohm Autolab, Bio-Logic, Gamry, Ivium Technologies, CHI, Pine Research, Admiral Instruments, etc.~to fulfill the same purpose, including models which provide wider current ranges, higher sample rates, and more measurement techniques (e.g. including impedance spectroscopy), the price of these instruments generally ranges from \$2000 up to \$20000 and more. To the author's best knowledge, the lowest-cost commercial instrument which could substitute for the presented design is the Squidstat Solo, sold by Admiral Instruments for a retail price of \$1900 \cite{squidstat}. While it has a higher sample rate (1~ms/sample, versus 90~ms for the presented design), a slightly higher potential range ($\pm$10V versus $\pm$8V) and similar current ranges ($\pm$3 $\upmu$A to $\pm$25 mA versus $\pm$2 $\upmu$A to $\pm$20 mA), it has worse potential and current resolution (16-bit versus 22-bit), it is approx.~20 times more expensive, and its hardware and software cannot be freely modified.

\section{The hardware}
\subsection{Circuit description}
\begin{figure*}[t]
 \centering
 \includegraphics[scale=1.1]{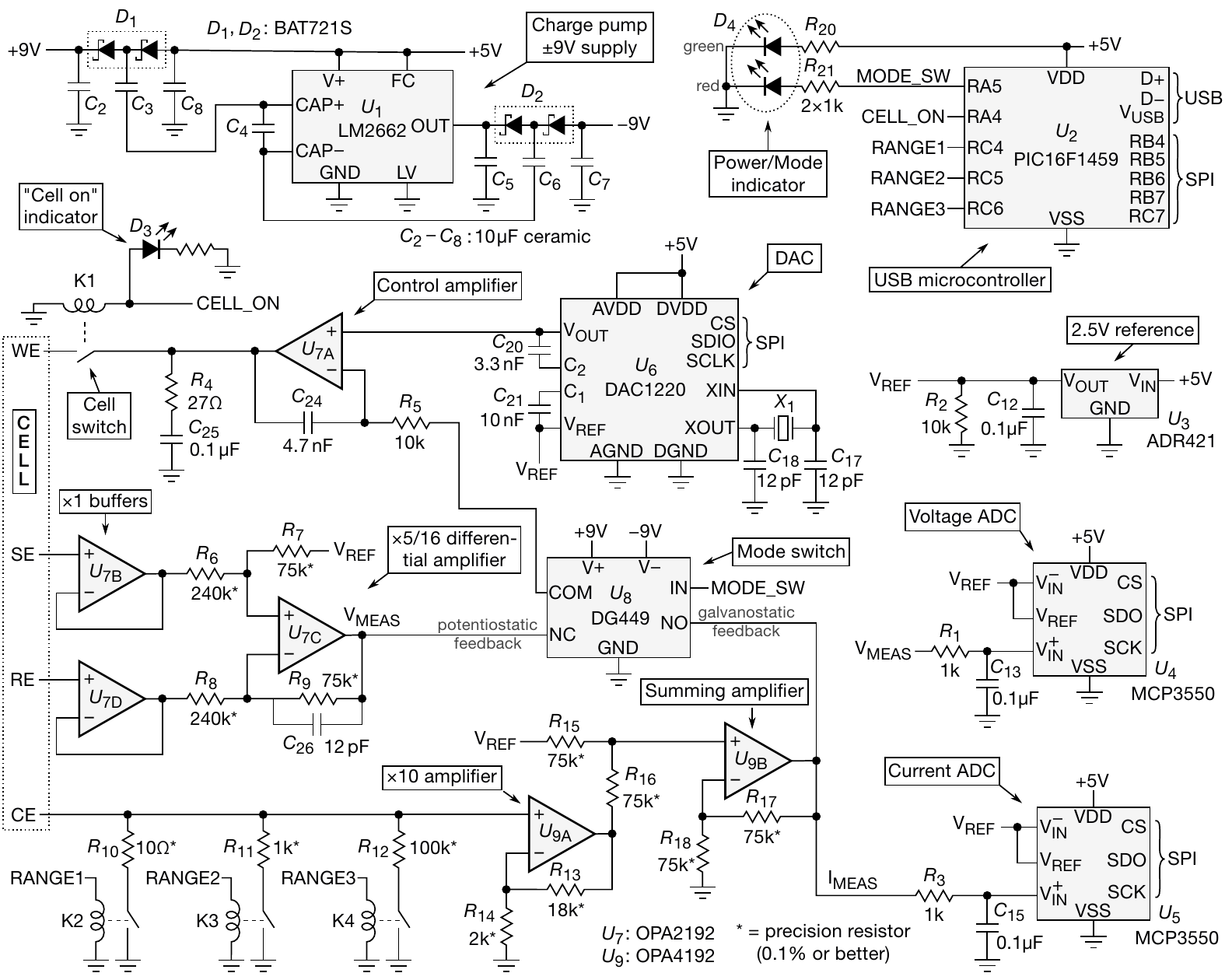}
 \caption{Annotated schematic diagram of the USB potentiostat/galvanostat circuit. Certain details are omitted for clarity; a complete and fabrication-ready circuit schematic can be found in the supplementary material \cite{sup}.}
 \label{fig3}
\end{figure*}

\noindent An annotated schematic diagram of the device is shown in Figure~\ref{fig3}. It consists of several subcircuits, which are discussed below.

\subsubsection{Power supply}
\noindent In order to be able to apply cell voltages between $-$8\,V and $+$8\,V, the analog circuitry needs dual power supply rails which supply at least these voltages (plus some overhead). The required current is fairly low; the cell current will not exceed 25~mA, and the quiescent current of the analog circuitry is in the order of a few mA. To eliminate the inconvenience of needing an external power supply, the supply is generated internally from the +5\,V line provided by the USB bus. This is achieved by a charge pump\footnote{The choice of a charge pump rather than a switched-mode power supply simplifies the design and eliminates inductors or transformers, which may emit undesirable electromagnetic noise.} consisting of U1 (an LM2662 switched-capacitor voltage convertor) and its associated circuitry. The switching action of U1 drives both a positive voltage doubler network (C2, C3, C4 and D1) and an inverting voltage doubler network (C5, C6, C7 and D2); this results in supply rails of $\pm$10\,V, minus the forward voltage losses of the diodes. These losses are minimized by choosing Schottky-type diodes such as the BAT721S, which conveniently houses two of them in a single package. The result is approx.~$\pm$9\,V. Ceramic capacitors are recommended for C2--C8; their low ESR results in low ripple, they do not degrade over time like electrolytics, and the required 10~$\upmu$F capacities are nowadays inexpensively available.

\subsubsection{Analog circuitry}
\noindent The analog circuitry is implemented using OPAx192 operational amplifiers. These relatively new op-amps from Texas Instruments~\cite{opax192} are high-precision, low offset voltage, low bias current, low-noise devices which have a wide supply voltage, rail-to-rail inputs and outputs, and a rather high output current. They closely approximate ideal op-amp behaviour, making them highly suitable for a measurement circuit like this where DC precision is of the utmost importance.

The core of the circuit is the control amplifier, U7A. It compares the voltage set by the DAC output to either the potentiostatic or the galvanostatic feedback voltage, and drives the working electrode of the cell until they are equal. To prevent oscillation, its bandwidth is limited by R5 and C24. The present component values yield a --3\,dB frequency of approx.~3~kHz, which is still much faster than the typical measurement timescale. An additional ``snubber''-type network consisting of R4 and C25 on its output pin increases stability towards capacitive loads. The cell switch K1 allows the electrochemical cell to be connected (this will later be referred to as the ``\texttt{CELL ON}'' state) or disconnected (``\texttt{CELL OFF}'') from the output of U7A; in its disconnected state, the cell is not driven, but may still be measured.

Potentiostatic feedback is acquired through U7B, U7C, and U7D. U7B and U7D are unity-gain buffers which present very high input impedances on respectively the sense electrode and the reference electrode, satisfying the requirement of having nearly zero current flow between SE and RE.\footnote{The OPAx192 has a typical input impedance of 10$^{13}\,\Upomega$ and a typical input bias current of 5~pA \cite{opax192}. Using e.g.~a reference electrode with an impedance of 10~k$\Upomega$, this results in an error voltage of 50~nV -- an insignificant quantity.} The buffered voltages are then fed into U7C, which is configured as a differential amplifier by means of R6--R9. This amplifier implements the following operation:
\begin{eqnarray*}
&V_{MEAS}=V_{REF}+(V_{SE}-V_{RE})\times(R7/R6) \\ &\text{(for $R8=R6$, $R9=R7$)}
\end{eqnarray*}

With $V_{REF}=\,$2.500\,V, $R7=$75.0\,k$\Upomega$ and $R6=$240\,k$\Upomega$, potential differences between SE and RE ranging from \mbox{--8\,V} to \mbox{+8\,V} are linearly scaled to output voltages $V_{MEAS}$ ranging from 0\,V to +5\,V. In this way, the signal spans the same range as the DAC output (allowing it to be used as a feedback signal) and the ADC input (allowing it to be measured).

Galvanostatic feedback is acquired by making use of the shunt resistors R10, R11 or R12 to convert the CE current into a voltage (selectable by the ranging relays K2--K4), multiplying this voltage by a factor of exactly 10 through the non-inverting amplifier U9A, and summing this voltage with $V_{REF}$ using U9B. In the highest current range, a range of --25~mA $\rightarrow$ +25~mA is mapped linearly to the range of 0 $\rightarrow$ +5V, which is again suitable as a feedback signal and for acquisition by the ADC. The lower current ranges are respectively 100 times and 10\,000 times more sensitive, resulting in ranges of resp.~$\pm$250~$\upmu$A and $\pm$2.5~$\upmu$A.

The potentiostatic and galvanostatic feedback signals (labelled \texttt{V\_MEAS} and \texttt{I\_MEAS}) lead into the ``normally closed'' and ``normally open'' terminals of U8, a DG449 analog switch. When \texttt{MODE\_SW} is low, the control amplifier receives the potentiostatic feedback signal; when high, it receives galvanostatic feedback. In this way, the circuit can quickly switch between the potentiostatic and galvanostatic control modes.

\subsubsection{A/D conversion}
\noindent The \texttt{V\_MEAS} and \texttt{I\_MEAS} signals are connected to U4 and U5, which are MCP3550 A/D convertors, for data acquisition. The MCP3550 is a 22-bit delta-sigma ADC which offers high accuracy and low noise; in particular, it strongly rejects line noise at either 50\,Hz (using the MCP3550-50 model) or 60\,Hz (using the MCP3550-60 model) \cite{mcp3550}. This is highly desirable because it eliminates what is often the most important noise source in many lab environments; however, this comes at the cost of a fairly limited conversion rate of max.~12.5 samples/s. If faster conversion would be required, it could be directly replaced by the MCP3553 which samples at 60~samples/s but lacks the line noise filter. The inputs to the ADCs are additionally low-pass filtered by R1/C13 and R3/C15 to remove high-frequency switching noise; the present component values yield roll-off frequencies of approx.~1.6~kHz, well below the delta-sigma modulator’s oversampling rate of 25600 samples/s (corresponding to a Nyquist frequency of 12.8 kHz) \cite{mcp3550}.

With 22-bit resolution, the potential is measured with a granularity of 3.8\,$\upmu$V. In the most sensitive current range, current is measured with a granularity of 1.2\,pA; this figure increases to resp.~120\,pA and 12\,nA for the higher ranges.

\subsubsection{D/A conversion and voltage reference}
\noindent The control amplifier receives its ``setpoint'' voltage from U6, a DAC1220 digital-to-analog convertor. The DAC1220 is a 20-bit delta-sigma DAC which is inherently linear and contains an on-chip calibration function \cite{dac1220}. It receives its +2.500\,V reference voltage $V_{REF}$ (as do U4, U5 and the analog circuitry) from U3, an ADR421 voltage reference using XFET technology for low noise, high accuracy ($\sim$0.1\%), and high stability \cite{adr421}.

The 20-bit DAC resolution results in potential control with a granularity of 15.3\,$\upmu$V, or in current control with granularities of 4.8\,pA, 480\,pA or 48\,nA, depending on the current range.

\begin{figure*}[t]
 \centering
\includegraphics[scale=1.12]{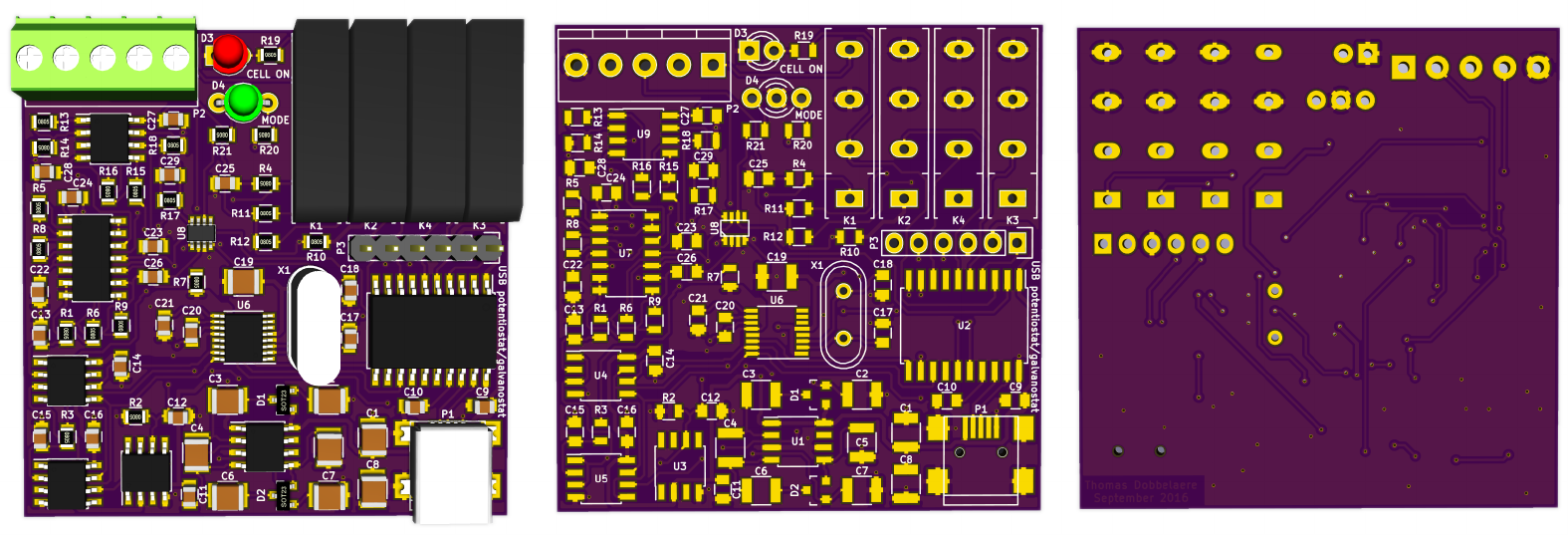}
  \caption{3D renders of the printed circuit board design, showing the populated board (left), the unpopulated board's front side (middle) and the unpopulated board's back side (right).}
  \label{fig4}
\end{figure*}

\subsubsection{Digital control}
\noindent The device connects to a host computer by an on-board USB interface. This function is implemented by U2, a PIC16F1459 microcontroller which has built-in USB capabilities and provides a sufficiently large number of general-purpose input/output pins for relay switching and SPI communication \cite{pic16f1459}. Its function is to receive commands from the host computer through USB, which may instruct it to either toggle a pin, read from the ADC, or set the DAC. It communicates with the ADC or DAC through a software-implemented SPI interface, and in case of a DAC read, it then sends back the acquired data to the host computer.

Status LEDs D3 and D4 provide some basic status indication. D3 provides the ``\texttt{CELL ON}'' indication; when D3 is illuminated, the cell is connected to the control amplifier. D4 is a dual-color LED which provides power-on and mode indication; it lights up green when the circuit is in the potentiostatic mode, and orange when it is in the galvanostatic mode.

\subsection{PCB design}
\noindent A compact, double-sided printed circuit board design was made in KiCad \cite{kicad}. The design files are available in the supplementary material \cite{sup}, and a 3D rendering of the (populated and unpopulated) board is shown in Figure~\ref{fig4}. The board is most easily fabricated by sending it off to a PCB prototyping service. Assembly can be carried out by either reflow soldering or hand-soldering.

\noindent Due to the small size of the PCB (approx.~5$\times$5\,cm), it may easily be put in a small enclosure, e.g.~a mint tin, provided some openings are made to allow the mini-USB connection and the cell connection cables to pass through.

\begin{figure}[b!]
 \centering
\includegraphics[scale=0.9]{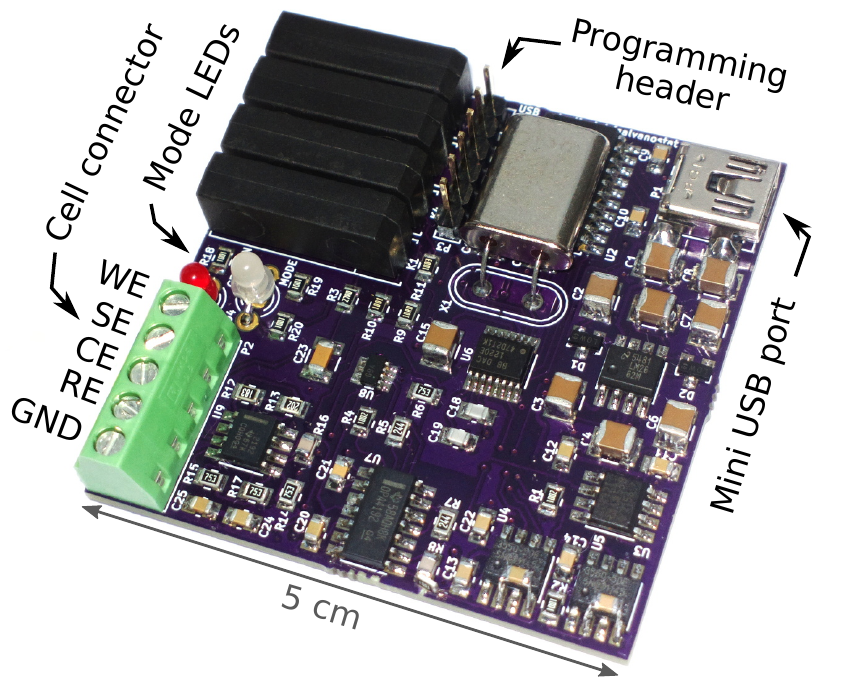}
  \caption{Annotated photograph of the assembled potentiostat.}
  \label{fig5}
\end{figure}

A bill of materials and a fabrication diagram (showing a top view of the PCB superimposed with the component values) may be found in the supplementary material \cite{sup} to aid component ordering and circuit assembly. The total cost of the device, including the components and a manufactured PCB, is well below \$100. A photograph of the finished device is shown in Figure~\ref{fig5}.

\subsection{Microcontroller firmware}
\noindent The microcontroller firmware including the source code and a compiled .hex image may be found in the supplementary material \cite{sup}. Compilation and usage instructions may be found in the sections below.

\subsubsection{Compilation}
\noindent The source code is written in the C programming language and can be compiled using the Microchip MPLAB XC8 compiler. The provided Makefile allows easy compilation because it automatically provides the compiler with the necessary flags and include paths. On a Linux system, simply running the command \texttt{make all} will compile and link everything and produce the output file \texttt{firmware.hex}.

\subsubsection{Raw USB communication}
\noindent The microcontroller firmware makes use of Signal 11's ``M-stack'' open-source USB stack \cite{mstack} to implement communication through raw USB bulk transfers. Commands are received as ASCII strings on EP1 OUT, are executed, and a reply is sent on EP1 IN. The following commands are supported:
\begin{itemize}
\item ``\texttt{CELL ON}'', ``\texttt{CELL OFF}'' \\ Connects or disconnects the cell to the output of the control amplifier.
\item ``\texttt{POTENTIOSTATIC}'', ``\texttt{GALVANOSTATIC}'' \\ Switches between potentiostatic or galvanostatic control modes.
\item ``\texttt{RANGE x}'' (where \texttt{x} = ``\texttt{1}'', ``\texttt{2}'', or ``\texttt{3}'') \\ Switches between current ranges; range 1 is the highest (least sensitive) current range, increasing range numbers yield lower (more sensitive) ranges.
\item ``\texttt{DACSET xxx}'' (\texttt{x} = one byte of data)\\ Sets the DAC output code (three bytes).
\item ``\texttt{DACCAL}'' \\ Performs an automatic DAC calibration.
\item ``\texttt{DACCALSET xxxxxx}'' (\texttt{x} = one byte of data)\\ Sets the DAC calibration data; the first three bytes are proportional to the offset, the latter three bytes are proportional to the gain \cite{dac1220}.
\item ``\texttt{DACCALGET}''\\ Returns six bytes, representing the current DAC calibration in the same format as above.
\item ``\texttt{ADCREAD}'' \\ Reads the potential and current from the ADCs, and returns six bytes if a previous conversion has finished. The first three bytes represent the potential ADC value, the latter three bytes represent the current ADC value. If a conversion is still ongoing, it returns the string ``\texttt{WAIT}'' instead.
\item ``\texttt{OFFSETSAVE xxxxxx}'' (\texttt{x} = one byte of data)\\ Saves six bytes to internal flash memory; these are used for potential/current offset removal.
\item ``\texttt{OFFSETREAD}'' \\ Retrieves the corresponding six bytes from internal flash memory and returns them.
\item ``\texttt{SHUNTCALSAVE xxxxxx}'' (\texttt{x} = one byte of data)\\ Saves six bytes to internal flash memory; these are used for fine-tuning the three current shunt resistors.
\item ``\texttt{SHUNTCALREAD}'' \\ Retrieves the corresponding six bytes from internal flash memory and returns them.
\end{itemize}
The host computer communicates with the potentiostat using a generic driver provided by the cross-platform libusb library \cite{libusb}.

\begin{figure*}[t]
  \centering
  \includegraphics[scale=1.05]{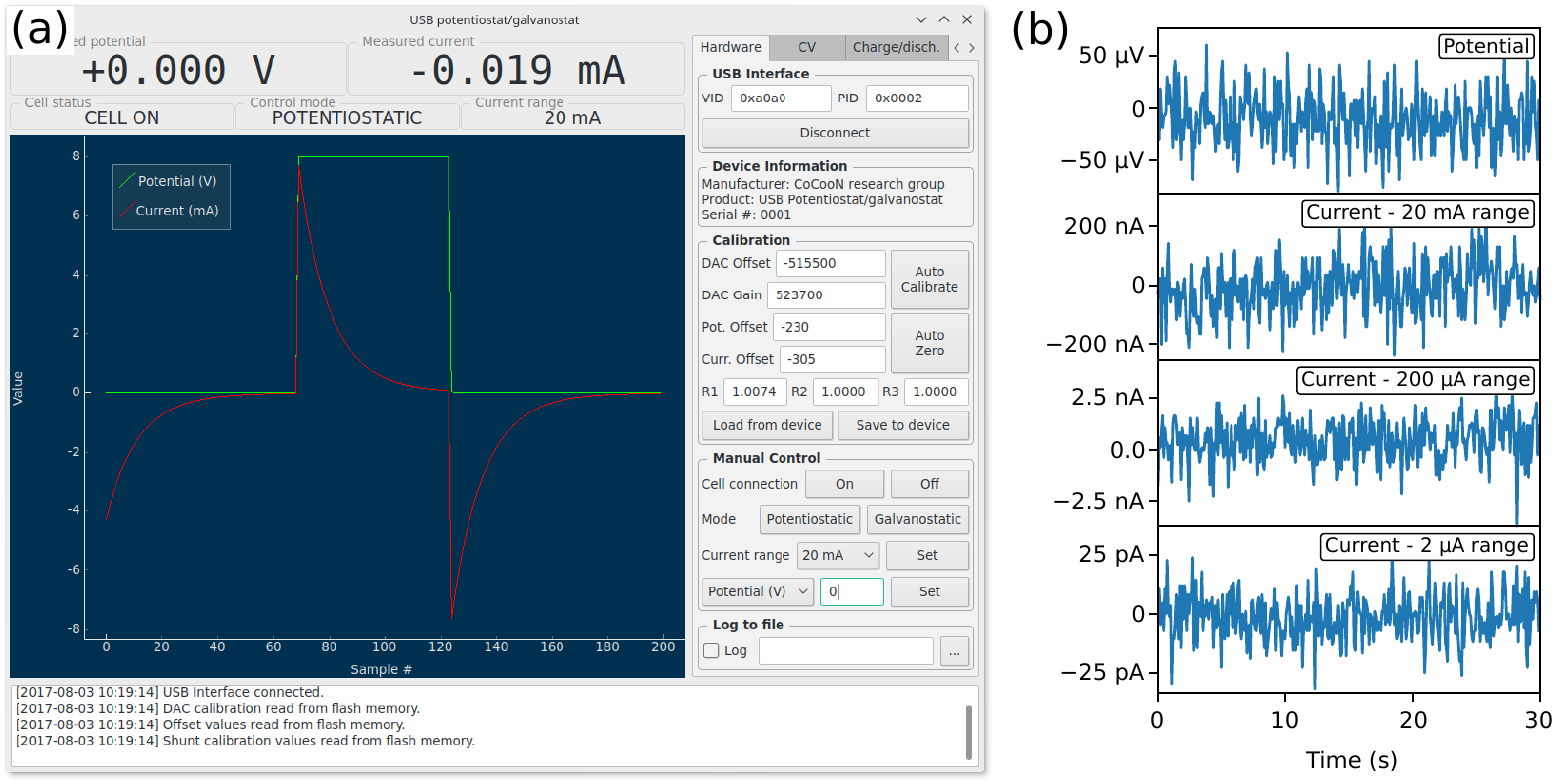}
  \caption{(a) The user interface program running on the host computer in idle mode, showing the measured potential and current in real time. (b) Noise levels on the measured potential (with shorted sense and reference electrodes, i.e.~zero potential) and on the measured current in the three current ranges (with the cell off, i.e.~zero current).}
  \label{fig6}
\end{figure*}

\section{The software}
\noindent A user-friendly GUI application allows the experimenter to easily perform typical electrochemical measurements. The program can be found in the supplementary material \cite{sup} and is written in Python 3. In addition to a working installation of Python, it requires NumPy \cite{numpy} and SciPy \cite{scipy} for data processing, PyUSB \cite{pyusb} for USB communication, PyQt \cite{pyqt} to provide the GUI, and PyQtGraph \cite{pyqtgraph} for real-time plotting. These packages are freely available, and may be easily installed from the system software repositories on most Linux distributions. On Windows and Mac, it is recommended to use Anaconda \cite{anaconda}; it provides Python~3.6 along with all packages except PyQtGraph (which may be installed using conda) and PyUSB (which may be separately installed using pip). Only on Windows, a USB device driver is required upon connecting the device to the host computer; this driver may be generated by libusb \cite{libusb} or downloaded from the supplementary material \cite{sup}.

After installing these dependencies, the program can be executed by running the command \texttt{python~tdstatv3.py}. An application GUI similar to the one in Figure~\ref{fig6}a should then appear. The USB Vendor and Product IDs shown in the input fields should match the values in the microcontroller firmware source code; the default values are resp.~\texttt{0xa0a0} and \texttt{0x0002}, but they can be adjusted if necessary. By clicking the ``Connect'' button, the application will start communicating with the potentiostat and will start displaying the measured potential and current; it does this both in numeric form (using the upper numeric indicators) and in graphical form (by plotting the potential and the current as a function of the time).

\subsection{Calibration}
\noindent Even though the device will already be reasonably accurate without applying any calibration (thanks to the precision op-amps, precision resistors, and the inherent linearity of the ADCs and DAC), the remaining -- small -- offset and gain errors, typically in the order of respectively 0.01\% of full-scale and 0.1\% of the measured value, can be calibrated out by adjusting the values in the ``Calibration'' field under the ``Hardware'' tab. It is recommended that this be done by the following procedure:
\begin{itemize}
\item Short the SE and RE; leave WE and CE unconnected. Either adjust the potential offset and current offset manually until both the potential and the current are exactly zero, or wait at least 20\,s (to have the device measure enough data points to average) and press the ``Auto Zero'' button.
\item Connect SE to WE, and RE to CE. Press the ``Auto Calibrate'' button. Under ``Manual Control'', switch the cell on, and set a few different potentials. The measured potential should exactly match the set potential; if it does not, you can further fine-tune the calibration by making small adjustments to the DAC offset and DAC gain. Press ``Save to device'' to apply the values.
\item Finally, the values R1--R3 represent relative fine adjustments to the shunt resistor values. If precise shunt resistors were used, they should not require adjustment, except for R1 (the lowest shunt resistor, used for the highest current range, having a nominal value of 10.00\,$\Upomega$) which may have a slightly higher value due to the non-negligible contact resistance of the current ranging relay. To fine-adjust this value, connect SE and We to one leg and WE and CE to the other leg of an accurate 1.000\,k$\Upomega$ resistor. Set a potential of e.g.~7.000\,V, set potentiostatic mode, and turn the cell connection on; adjust the R1 value until the measured current is exactly 7.000~mA.
\item By pressing the ``Save to device'' button, the calibration data is saved to the internal flash memory of the potentiostat, so that the correct values are automatically loaded upon the next application start-up. This is also useful in case the potentiostat is connected to a different host computer; the device will remember its own calibration settings.
\end{itemize}

To maintain the best accuracy, it is recommended to verify the calibration or repeat the calibration procedure if the device has been exposed to significant changes in ambient temperature or humidity.

\subsection{Measurement}
\begin{figure}[h!]
 \centering
\includegraphics[width=85mm]{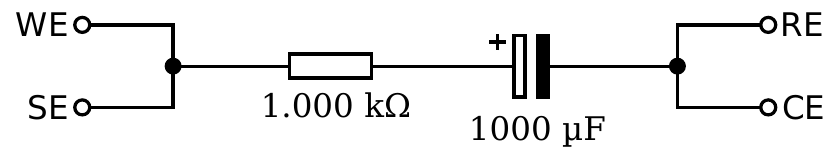}
  \caption{Schematic diagram of the dummy cell.}
  \label{fig7}
\end{figure}

\begin{figure}[b!]
 \centering
\includegraphics[width=87mm]{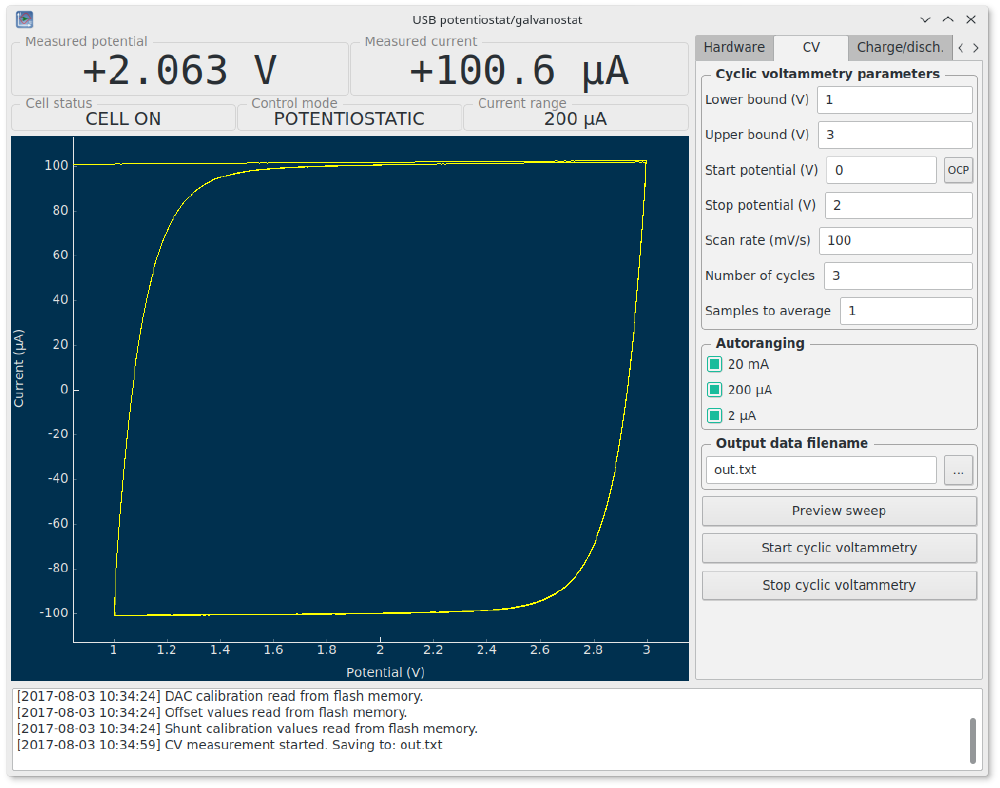}
  \caption{Running a cyclic voltammetry measurement on the dummy cell using the ``CV'' tab.}
  \label{fig8}
\end{figure}

\noindent The user interface program can run a number of electrochemical techniques, which can be started from within the corresponding tab views. Figures~\ref{fig8}--\ref{fig12} illustrate these techniques using example measurements performed on a ``dummy cell''. The dummy cell consists of a series circuit of a 1.000\,k$\Upomega$ precision resistor and a 1000\,$\upmu$F (nominal) electrolytic capacitor, connected on one side to WE/SE and on the other side to RE/CE, and is drawn schematically in Figure~\ref{fig7}. The RC series circuit can be used to verify the correct operation of the potentiostat by comparing the measured potential and current to the expected behavior, which is governed by a first-order differential equation:
\begin{eqnarray}
I=&\frac{U_R}{R}=C\frac{dU_c}{dt} \\ U=&U_R+U_C
\end{eqnarray}

\subsubsection{Constant potential / current}
\noindent Using the manual controls in the ``Hardware'' tab, a fixed potential (in potentiostatic mode) or a fixed current (in galvanostatic mode) may be set. This makes it possible to use the device as a constant-voltage or constant-current source. Both the potential and current will be continuously measured and displayed. It is possible to log the data to a file by entering a filename and checking the ``Log'' checkbox. Figure~\ref{fig6}a shows the result of periodically changing the potential over the dummy cell between 0\,V and +8\,V (green curve), resulting in current spikes which decay exponentially as expected for an RC circuit (red curve). Figure~\ref{fig6}b shows the noise on the measured potential and current, using shorted leads (i.e.~zero potential) and with the cell connection off (i.e.~zero current). The RMS noise level on the potential is 28~$\upmu$V, while the RMS noise levels on the current are respectively 88~nA, 1.1~nA, and 9.9~pA in the 20~mA, 200~$\upmu$A, and 2~$\upmu$A ranges.

\subsubsection{Cyclic voltammetry}
\noindent Cyclic voltammetry (CV) is a versatile electrochemical measurement technique which finds application in many fields; the details can be found in literature \cite{bard_faulkner}, but it is essentially based on the repeated application of a linear potential ramp, i.e.~a ``triangle waveform'', while measuring the resulting current. In a typical CV plot, the current is then plotted as a function of the applied potential.

\begin{figure}[h!]
 \centering
\includegraphics[width=85mm]{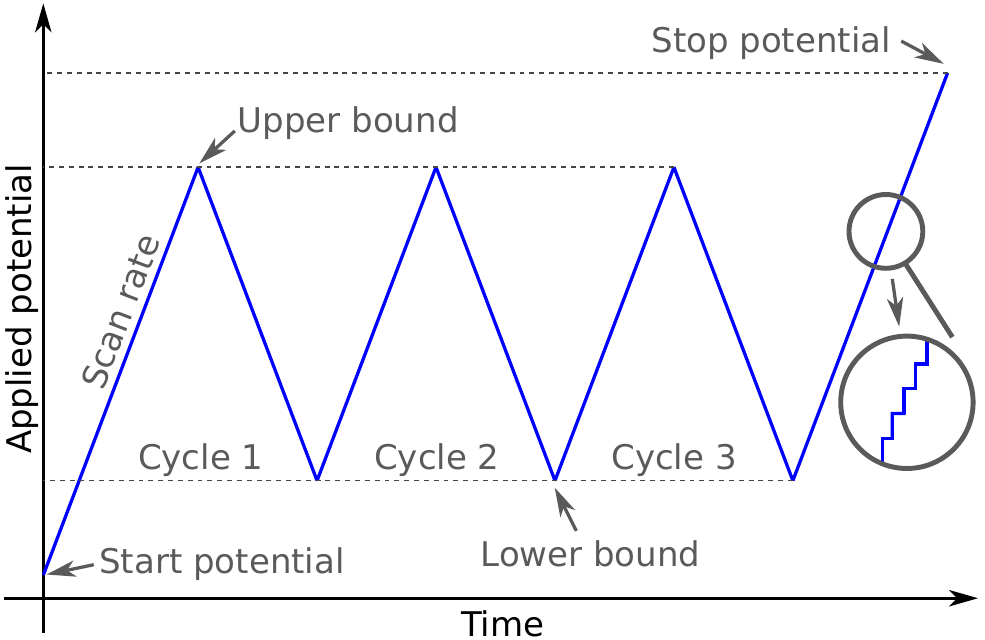}
  \caption{Graphical illustration showing the applied potential waveform during a cyclic voltammetry measurement, and the parameters that determine its shape.}
  \label{fig9}
\end{figure}

Before starting a CV measurement, a number of parameters must be set in the user interface program (Figure~\ref{fig8}). These parameters determine the nature of the applied triangular waveform and are graphically illustrated in Figure~\ref{fig9}. Because of the finite resolution (i.e.~step size) in potential and time, the waveform is actually not truly linear but rather consists of a ``staircase'' shape; however, when the steps are sufficiently small, the results are equivalent to a truly linear ramp.

\noindent By clicking the ``OCP'' button next to the ``Start potential'' input field, the currently measured potential is copied into the input field; this is convenient when starting a CV measurement at the open-circuit potential (OCP) of the cell.

The ``Samples to average'' parameter is automatically calculated based on the scan rate, but can be overwritten. It determines how many samples are averaged for one measurement of potential and current. Samples are acquired every 90~ms; this means that a measurement containing $n$ averaged samples will take $n\,\times\,$90~ms. Averaging reduces noise, but also reduces the effective sampling rate; thus, $n>1$ should only be used for sufficiently slow scan rates. Because the minimum sampling time is 90~ms, the maximum scan rate is limited to approx.~500~mV/s; at this scan rate, the height of the potential steps is 45~mV.

Because the resulting current is not known \emph{a priori} and may span a large dynamic range, an autoranging feature is available: during the measurement, the device will automatically choose an appropriate current range based on the actually measured current. This allows it to accurately measure currents ranging from nanoamps to max.~25~mA. If this is, for some reason, not desirable, a current range can be ``disabled'' (i.e.~prevented from being selected by autoranging) by unticking its checkbox.

Before starting the measurement, the ``Preview sweep'' button may be used to make a potential vs.~time plot based on the currently set CV parameters. This allows the experimenter to verify the CV settings. After setting an output filename, the measurement can be started by clicking ``Start cyclic voltammetry''. This will switch the device to potentiostatic mode and start applying the potential profile. During the measurement, the application shows a continuously updating CV plot and continuously writes the measurement data to the output file. The output file is formatted as ASCII text and contains three tab-separated columns representing resp.~the elapsed time (in s), the measured potential (in V), and the measured current (in A). It may be imported in any plotting or data analysis tool.

The CV curve of the dummy cell (Figure~\ref{fig8}), acquired at a scan rate of 100~mV/s, reveals nearly horizontal plateaus at currents of $\pm$100.6\,$\upmu$A (with the plus sign on the rising potential and the minus sign on the declining potential). This is expected; because the current through the capacitor is proportional to the rate of change of its voltage, and that (given a constant current) the voltage drop over the resistor is constant, the rate of change of the capacitor voltage is also equal to $\pm$100\,mV/s. A straightforward calculation reveals that a 1006\,$\upmu$F capacitance yields the measured constant current. This value is well within the tolerance of the (nominally) 1000\,$\upmu$F capacitor in the dummy cell.

\begin{figure}[b!]
 \centering
\includegraphics[width=87mm]{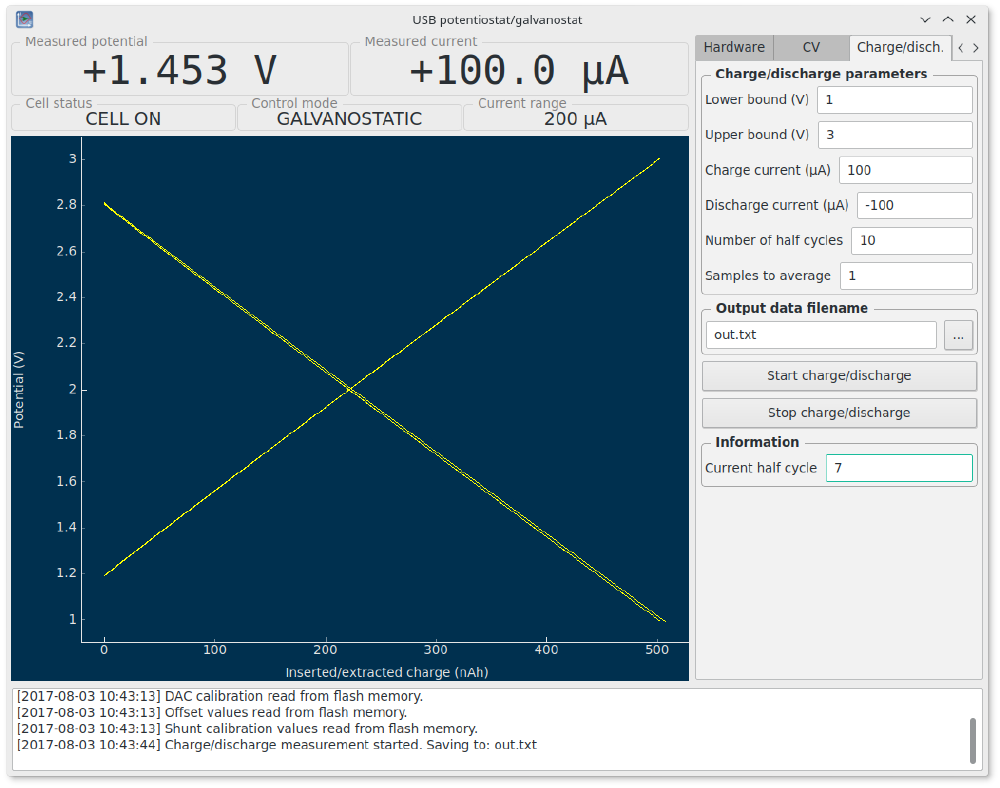}
  \caption{Running a charge/discharge measurement on the dummy cell using the ``Charge/Disch.'' tab.}
  \label{fig10}
\end{figure}

\subsubsection{Constant-current charge/discharge}
Another commonly used electrochemical technique is to apply a constant current and to observe the evolution of the measured potential over time. In analytical redox chemistry, this method is known as coulometric titration. In the field of battery research, it can be used to determine the capacity of an electrode material. Using the galvanostatic mode of the potentiostat, such a measurement technique is easily implemented. Figure~\ref{fig10} shows a charge/discharge measurement running on the dummy cell. Before starting a measurement, the following parameters need to be set in the ``Charge/Disch.'' tab:
\begin{itemize}
\item Upper bound and lower bound: during the charging phase, the charge current is applied until the measured potential reaches the upper bound. This marks the end of the charging phase and the beginning of the discharging phase. During the discharging phase, the discharge current is applied until the measured potential reaches the lower bound.  This marks the end of the discharging phase and the beginning of the next charging phase, thus repeating the cycle.
\item Charge and discharge current: sets the applied currents (in $\upmu$A) during the charge, resp.~discharge phases.
\item Number of half cycles: sets the total number of charge and discharge phases. As opposed to counting full cycles, this allows the experimenter to carry out e.g.~a single charge measurement (one half cycle), or a charge/discharge/charge measurement (three half cycles).
\item Samples to average: has the same function as explained earlier for cyclic voltammetry. Set it to the desired acquisition period, divided by 90~ms; higher values reduce noise, but result in slower data acquisition.
\end{itemize}

\begin{figure}[h!]
 \centering
\includegraphics[width=85mm]{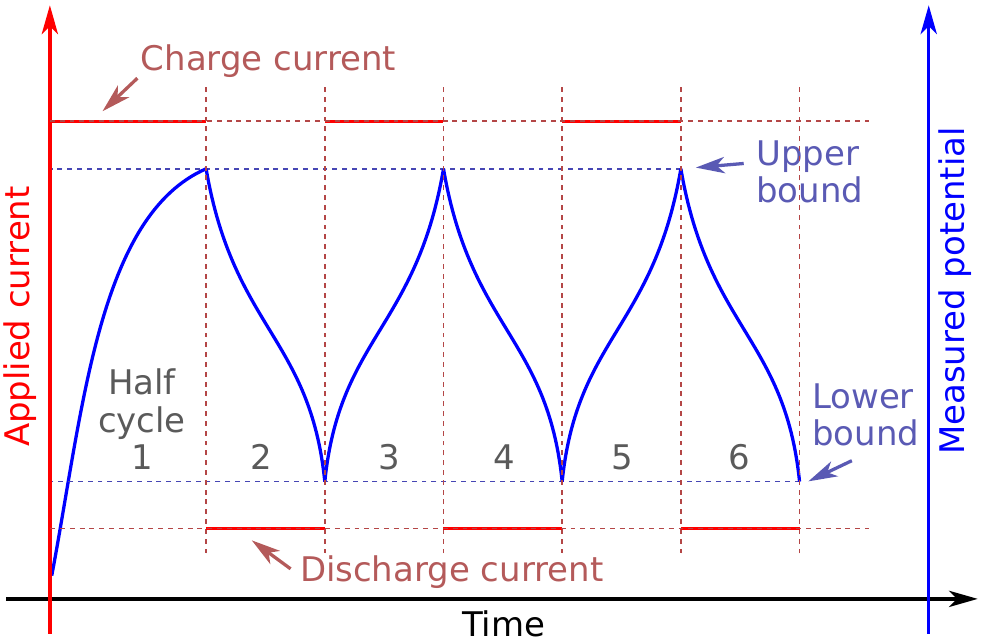}
  \caption{Graphical illustration of the applied current (in red) during a charge/discharge measurement, assuming a fictive potential profile (in blue).}
  \label{fig11}
\end{figure}

The meaning of these parameters in the context of a typical charge/discharge measurement is illustrated graphically in Figure~\ref{fig11}.

After setting the measurement parameters and choosing an output filename, clicking the ``Start charge/discharge'' button will start the charge/discharge process. During the measurement, the current half cycle is indicated in the ``Information'' box, and the plot window shows a continuously updating plot of the potential vs.~the charge. The charge is calculated as the absolute value of the product of the (constant) current and the elapsed time.

For the dummy cell, the resulting measurement is shown in Figure~\ref{fig10}. In the charge phase, using a constant current $I$ of +100\,$\upmu$A, the potential over the resistor $U_R$ has a constant value of +100\,mV, and the potential over the capacitor $U_C$ increases linearly at a rate of $dU_C/dt=100$~mV/s. This corresponds to the observed behavior. In the discharge phase, the same current is applied in the opposite direction. This causes the potential to decrease linearly at the same rate. The inserted/extracted charge over the 1.8\,V potential window equals 100\,$\upmu$A$\times$(1.8\,V/0.1\,V/s) = 1.8 mC = 500 nAh, which is in agreement with the value indicated on the horizontal axis.

\subsubsection{Rate testing}
Specifically in battery research, the experimenter is often interested in the ``rate behavior'' of a test cell. This refers to the influence of the charge/discharge current on the measured cell capacity; typically, the capacity decreases with increasing current due to kinetic limitations. The current is expressed as a ``C-rate'', where 1C is defined as the current necessary to charge/discharge the cell to its theoretical capacity in exactly one hour.

\begin{figure}[h!]
 \centering
\includegraphics[width=87mm]{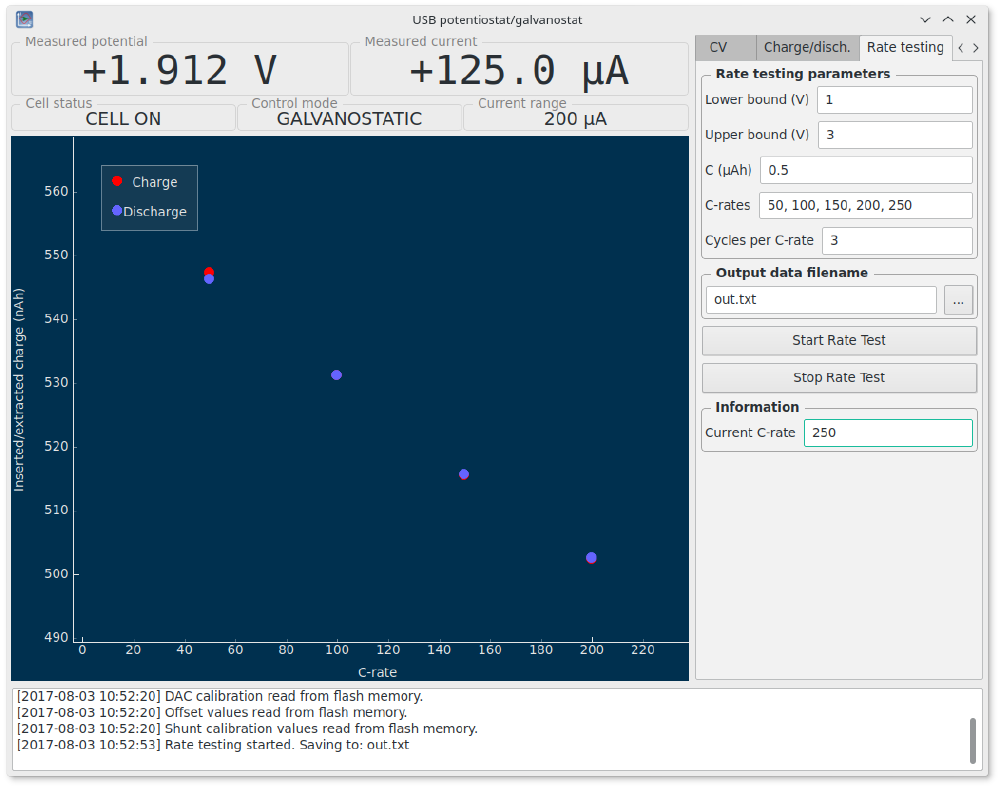}
  \caption{Running a rate testing measurement on the dummy cell using the ``Rate testing'' tab.}
  \label{fig12}
\end{figure}

Using the ``Rate testing'' feature which is accessible from the correspondingly named tab in the user interface program (see Figure~\ref{fig12}), the measurement of this rate behavior can be automated. For each C-rate (multiple values may be separated by commas), a charge/discharge measurement is ran between the lower and upper potential bounds (as explained in the previous section) for a configurable number of cycles. The charge/discharge current is calculated as:
\begin{equation}
I\,[\upmu\text{A}]=\pm C\,[\upmu\text{Ah}]\times\text{C-rate}
\end{equation}
with the positive sign for the charge current, and the negative sign for the discharge current.

During the measurement, the plot continuously updates itself to show the charge/discharge capacity measured in the final cycle as a function of the C-rate. The measurement on the dummy cell (Figure~\ref{fig12}) reveals a capacity which decreases linearly with the C-rate; indeed, the potential that is ``lost'' over the resistor is proportional to the applied current. The fact that the capacity decreases with the current, just like for a real battery, is no coincidence; in fact, the resistor in the dummy cell can be considered to represent the internal resistance encountered in real batteries.

\section{Use case: a thin-film lithium-ion battery electrode}
In the following section, the utility of this low-cost potentiostat will be demonstrated by an example use case in the context of thin-film lithium-ion batteries. A lithium-ion test cell was constructed using a PTFE body filled with electrolyte (1\,M \ce{LiClO4} in propylene carbonate) which was clamped against a 40~nm anatase \ce{TiO2} film on a TiN-coated silicon substrate, giving an active area of 0.95~cm$^2$. An electrical contact was made by applying conductive silver ink on the cleaved sides of the substrate in order to connect the TiN current collector layer to a piece of copper foil. This formed the working electrode. The counter electrode and reference electrode consisted of lithium metal strips dipped into the electrolyte.

As lithium-ion cells are incompatible with moisture and oxygen, the whole cell was constructed inside an argon glovebox. Thanks to its small size, the potentiostat could also be put inside the glovebox, needing only a USB cable feedthrough. Moving the potentiostat close to the test cell (inside the glovebox) greatly reduces noise and interference, compared to having it outside and extending the cell connection leads.

\begin{figure}[h!]
 \centering
\includegraphics[width=85mm]{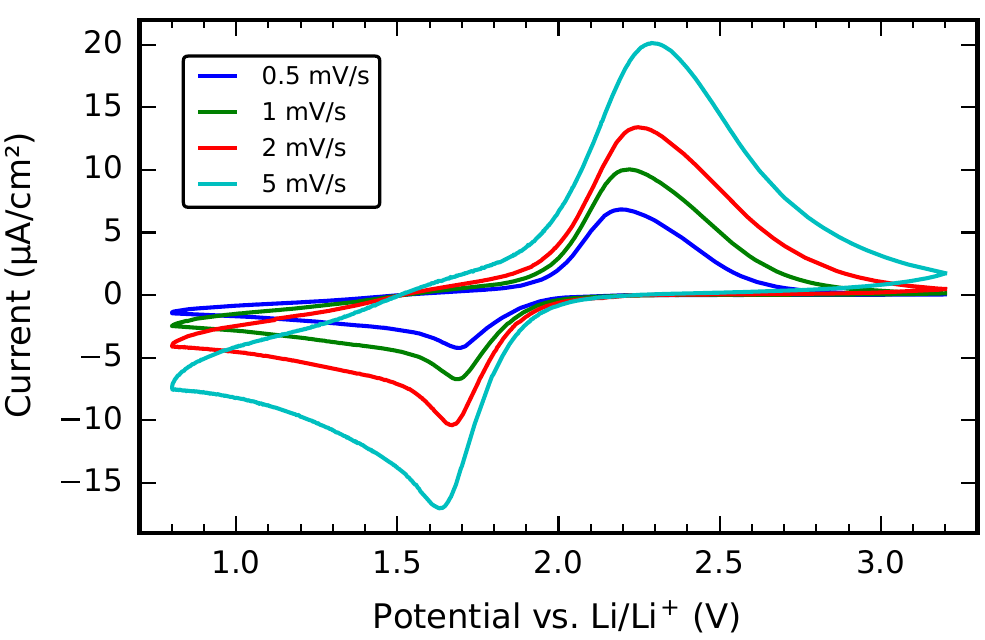}
  \caption{Cyclic voltammogram of a 40~nm \ce{TiO2} electrode acquired between 0.8\,V and 3.2\,V at scan rates between 0.5~mV/s and 5~mV/s.}
  \label{fig13}
\end{figure}

Cyclic voltammetry patterns were acquired at scan rates between 0.5~mV/s and 5~mV/s and are shown in Figure~\ref{fig13}. The cathodic and anodic peaks are clearly visible around resp.~1.7\,V and 2.2\,V, as is expected for anatase \ce{TiO2}.\cite{wang2007} As the scan rate increases, the available time for charge transfer decreases, causing the peak current to increase. The separation between the cathodic and anodic peak currents also increases; this is a kinetic effect, caused by the limited speed at which the lithium ions can diffuse through the \ce{TiO2} film. This makes such a CV experiment useful to study the electrode kinetics. A thorough study can already be found in literature~\cite{wang2007} and will not be repeated here.

\begin{figure}[h!]
 \centering
\includegraphics[width=85mm]{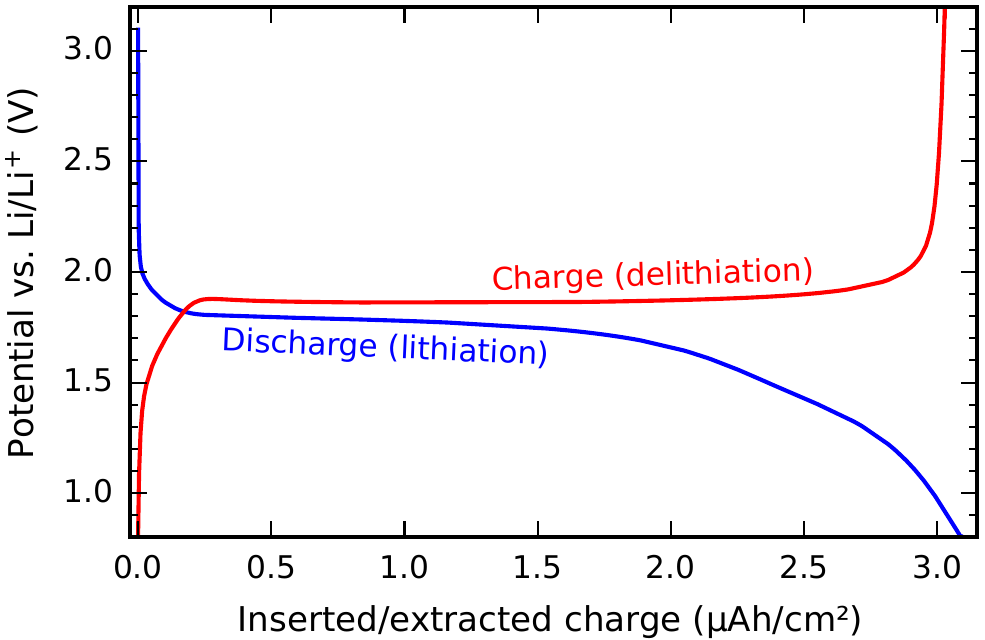}
  \caption{Potential evolution during constant-current charging and discharging of a 40~nm \ce{TiO2} electrode at a current density of $\pm$0.18~$\upmu$A/cm$^2$.}
  \label{fig14}
\end{figure}

The same electrode was subjected to charge/discharge cycling at a constant current of $\pm$0.18~$\upmu$A/cm$^2$ (positive sign for charge, negative for discharge) and between the same potential limits as in the CV experiment. The potential evolution, plotted as a function of the inserted/extracted charge, is shown in Figure~\ref{fig14}. It shows characteristic plateaus at 1.85\,V (charge) and 1.75\,V (discharge). The final capacities are 3.03~$\upmu$Ah/cm$^2$ (charge) and 3.09~$\upmu$Ah/cm$^2$ (discharge), yielding a coulombic efficiency (defined as the ratio between the delithiation and the lithiation capacities) of 98.1\%. The measured capacity corresponds to the insertion/extraction of approx.~0.6~Li$^+$ per unit of \ce{TiO2}, according to the following calculation:
\begin{equation}
\#\text{Li}=\frac{C_\text{measured}\times M}{d \rho N_A q_e}
\end{equation}
\\[0pt]
with $C_\text{measured}\approx0.011$~C/cm$^2$, $d=40\times10^{-7}$~cm, $\rho=3.8$~g/cm$^3$, $M=79.9$~g/mol, $N_A$ as Avogadro's constant, and $q_e$ as the elementary electron charge. 

\begin{figure}[h!]
 \centering
\includegraphics[width=85mm]{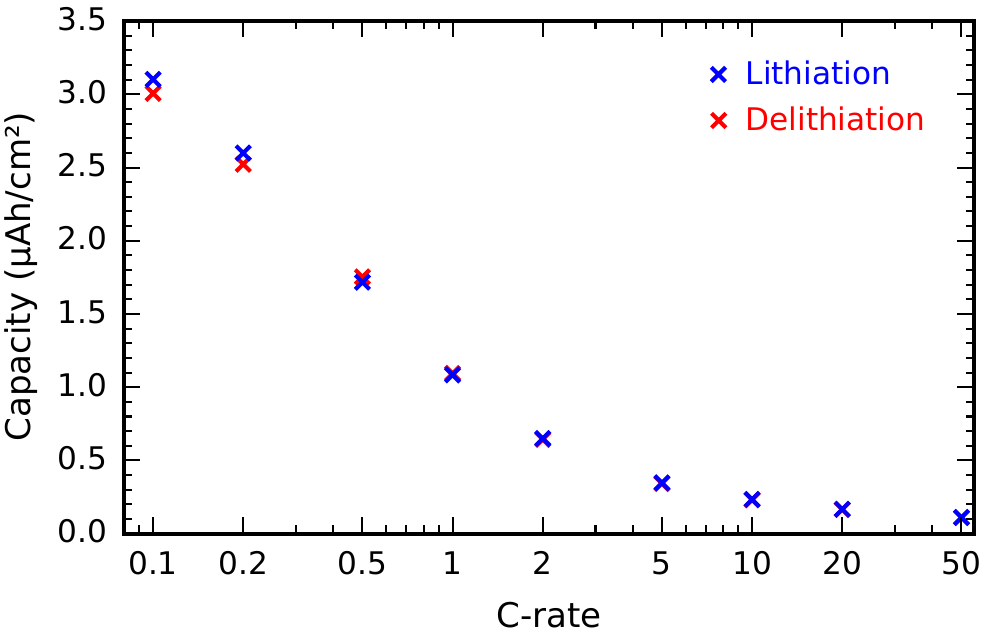}
  \caption{Rate testing results of the 40~nm \ce{TiO2} electrode, showing the measured capacity as a function of the C-rate.}
  \label{fig15}
\end{figure}

Taking the average of the measured charge and discharge capacity and defining 1C as 3.06~$\upmu$Ah/cm$^2$, the current density of $\pm$0.18~$\upmu$A/cm$^2$ can be equivalently expressed as a C-rate of approx. C/17, which is a rather slow rate. The charge/discharge measurements were repeated with increasingly higher currents using the rate testing mode of the potentiostat. The capacity was determined for each C-rate and the result is shown in Figure~\ref{fig15}. The results indicate that the capacity decreases quickly with an increasing C-rate, demonstrating the sluggish kinetics of the \ce{TiO2} electrode. At a rate of C/2, only approx.~50\% of the C/10 capacity remains. At very high C-rates ($\geq$10C), the slope of the decline becomes flatter; in this region, only a small pseudo-capacitive contribution to the capacity remains.

\section{Limitations and possible modifications}
Although this potentiostat design may be used ``as-is'' for many applications, some may have requirements that exceed its capabilities in its current state. Because the design is fully open source, it is entirely possible to modify it in order to accommodate custom requirements. In particular:
\begin{itemize}
\item In the present design, the maximum current of 25~mA is limited by (1) the current that can be delivered by the charge pump, and (2) the maximum output current of the control amplifier U7A. A higher current could be achieved by (1) replacing the charge pump by an external $\pm$9\,V power supply (which must be able to supply the desired current), (2) including a high-current buffer stage (e.g.~a pair of medium current transistors in a class AB amplifier configuration), and (3) adding a low-value current shunt resistor (e.g.~1$\Upomega$ for a maximum current of 250~mA) and ranging relay.
\item The sampling period of 90~ms may be too slow in case of fast electrochemical processes. It can be straighforwardly lowered to approx.~17~ms by replacing the MCP3550-type ADCs with the MCP3553 and adjusting the polling interval in the Python software. Even faster sampling would require more elaborate modifications.
\item A new experimental technique may be required. As these are implemented in the software running on the host computer, it is relatively straightforward to add a new technique by editing the Python source code.
\end{itemize}

\begin{acknowledgments}
T. Dobbelaere thanks the Fund for Scientific Research - Flanders (FWO) for financial support.
\end{acknowledgments}

\end{document}